\documentclass[12pt]{article}
\def\beq{\begin{equation}}
\def\eeq{\end{equation}}
\def\bea{\begin{eqnarray}}
\def\eea{\end{eqnarray}}
\def\bq{\begin{quote}}
\def\eq{\end{quote}}

\def\gappeq{\mathrel{\rlap {\raise.5ex\hbox{$>$}}
{\lower.5ex\hbox{$\sim$}}}}

\def\lappeq{\mathrel{\rlap{\raise.5ex\hbox{$<$}}
{\lower.5ex\hbox{$\sim$}}}}

\begin{document}
\topmargin -0.5cm
\oddsidemargin -0.3cm
\pagestyle{empty}
\begin{flushright}
{HIP-2008-34/TH}
\end{flushright}
\vspace*{5mm}
\begin{center}
\large
{\bf On non-holonomic systems and variational principles} \\
\normalsize
\vspace*{1.5cm} 
{\bf Christofer Cronstr\"{o}m}$^{*)}$ \\
\vspace{0.3cm}
Helsinki Institute of Physics\\
P. O. Box 64, FIN-00014 University of Helsinki, Finland \\
\vspace{0.3cm}
and\\
\vspace{0.3cm}
{\bf Tommi Raita}$^{**)}$\\
\vspace{0.3 cm}
Physics Department\\
P. O. Box 64, FIN-00014 University of Helsinki, Finland \\
\vspace{0.3cm}
 
\vspace*{1cm}

{\bf ABSTRACT} \\

\end{center}

We consider the compatibility of the equations of motion which follow from d'Alembert's principle in the case of a general autonomous non-holonomic mechanical system in $N$ dimensions, with those equations which follow for the same system by assuming the validity of  a specific variational action principle, in which  the non-holonomic conditions are implemented by means of the multiplication rule in the calculus of variations.  The equations of motion which follow from the principle of d'Alembert are not identical to the equations which follow from the variational action principle. We give a  proof that the solutions to the equations of motion which follow from d'Alembert's principle  do not  in general satisfy the equations which follow from the action principle with non-holonomic constraints. Thus the principle of d'Alembert and the minimal action principle involving the multiplication rule are not compatible in the case of systems with non-holonomic  constraints. For simplicity the proof is given for autonomous systems only, with one general non-holonomic constraint, which is  linear in the generalized velocities of the system. 
\vspace*{5mm}
\noindent

\vspace*{1cm} 
\noindent 

\noindent 
$^{*)}$ e-mail address: Christofer.Cronstrom@Helsinki.fi \\ 
$^{**)}$ e-mail address: Tommi.Raita@Helsinki.fi

\vfill\eject

\section{Introduction}
\pagestyle{plain}
\pagenumbering{arabic}
\setcounter{page}{1}

Hamilton's principle for mechanical systems with non-holonomic constraints has  recently been discussed by Flannery \cite{Flannery}.  In particular a variational formulation of the equations of motion of a mechanical system was discussed both for holonomic and non-holonomic constraints. It was shown  that while the equations of motion for a system with holonomic constraints can be obtained as variational equations,  with the  constraints being taken into account by the multiplication rule in the calculus of variations \cite{Mikhlin},  the corresponding procedure with non-holonomic constraints leads to equations which differ  from the equations of motion which follow from the well-known principle of d'Alembert.  

The problems discussed by Flannery are not new;  they have been discussed in the literature at least since Hertz's textbook \cite{Hertz},  in which the use of variational principles in mechanics was questioned. We refer in particular to an early paper by H{\"o}lder \cite{Holder}, in which  essential differences between holonomic and non-holonomic systems was discussed. Two later papers published  by Jeffreys \cite{Jeffreys} and Pars \cite{Pars} considered again Hamilton's principle for non-holonomic systems, and proposed  rectification  of previous papers in which the variational procedure (action principle) involving the multiplication rule had been proposed for systems with non-holonomic constraints.

The papers by Flannery, Jeffreys and Pars referred to above, contain several references to  papers  in which one has advocated  the use of a variational action principle  involving the multiplication rule for non-holonomic systems.  We should also mention a paper by Berezin \cite{Berezin}, in which it is taken for granted that  the same minimal action principle is valid both for holonomic and non-holonomic systems, with the constraints taken into account using the multiplication rule in the calculus of variations.

Even though the equations of motion following from the principle of d'Alembert and from the variational action principle with non-holonomic constraints are different in form, one may still argue that the equations in question may have the same solutions. It was demonstrated by Pars \cite{Pars}, that  the solutions to the equations in question differ from one another, in the case of a simple example of a non-holonomic system  in three dimensional space. As such, this validates the assertion that the solutions to the two different types of equations of motion are different in general, at least in a space of three dimensions.  To the best of our knowledge it has not been proved for general systems that the the two kinds of equations of motion referred to above in general have different solutions. We will supply a proof of this fact in this paper, for a general autonomous system with a finite number of degrees of freedom, restricted only by reasonable smoothness conditions.  For simplicity we consider in our proof only the case of one non-holonomic  constraint, which is taken to be linear and homogeneous in the  generalized velocities of the system. 

The problem we consider ought to be of interest in Classical Mechanics, but may also be of interest for the question of quantization  of non-holonomic systems.  One standard method of quantization of a classical system, namely the functional integral method, relies explicitly on the existence of an  action principle for the classical system in question. This is also true for the case of canonical quantization, which requires the existence of canonical coordinates and momenta and a proper Hamiltonian. This, in turn, requirers the existence of a proper Lagrangian, and hence the existence of an action principle. 

The proof given in this paper implies definitely that the standard action principle, which is valid for unconstrained systems, and which can be  generalized to cover the case of holonomic constraints by using the multiplication rule in the calculus of variations, is not consistent with the principle of d'Alembert if  the  non-holonomic constraints are implemented by means of the multiplication rule. Hence this action principle can not be taken as a starting point for quantization of non-holonomic systems, unless one dismisses the principle of d'Alembert in the case of non-holonomic systems, and assumes the validity of the action principle involving the multiplication rule instead.  

In the next section we consider the  Lagrange equations of motion which follow from the principle of d'Alembert for an autonomous mechanical system with both holonomic and non-holonomic constraints. 
We also derive the corresponding variational equations, which are obtained by using the multiplication rule for the implementation of the constraints.


\section{Lagrange equations with  constraints}

Consider an autonomous  mechanical system with independent generalized coordinates 
$q = (q^{1},...,q^{N})$, and  velocities $\dot{q} = (\dot{q}^{1},...,\dot{q}^{N})$. We denote the kinetic energy of the system  by  $T$, and the generalized applied forces on the system by  $Q_{A}, A = 1,...,N$. The principle of d'Alembert (see  e.g.  the classical texts by Goldstein \cite{Goldstein59}  or Whittaker \cite{Whittaker}) then gives the following equation,
\beq
\label{Lag0}
\sum_{A=1}^{N} \left \{\frac{d}{dt}\left (\frac{\partial T}{\partial  \dot{q}^{A}} \right ) - \frac{\partial T}{\partial q^{A}} - Q_{A}\right \} \delta q^{A} = 0,
\eeq
where the quantities $\delta q^{A}$ are  {\em virtual displacements} of the system. If the virtual displacements $\delta q^{A}, A = 1,\ldots, N,$ are independent, then the Eqn.\ (\ref{Lag0}) results in the Lagrange equations of motion,
\beq
\label{Lag1}
\frac{d}{dt}\left (\frac{\partial T}{\partial  \dot{q}^{A}} \right ) - \frac{\partial T}{\partial q^{A}} = Q_{A}, \; A = 1, \ldots, N.
\eeq

We  generalize  to systems with $1 \leq m  <  N$ independent non-holonomic constraints, which are taken to be linear and homogeneous in the velocities. The non-holonomic constraints  are thus of the following form,
\beq
\label{nholco}
\sum_{A=1}^{N} a^{i}_{A}(q)  \dot{q}^{A} = 0,\,  i = 1,..., m < N,
\eeq
where the quantities $a^{i}_{A}(q)$  are given functions of the variables $q = (q^{1},...,q^{N})$. 

The derivation given below of the equations of motion for this non-holonomic system can be found  in the textbook by  Whittaker \cite{Whittaker}.


Implement the constraints (\ref{nholco}) by regarding the system to be acted on by both the external applied forces  $Q_{A}$ and  by certain additional forces of constraint $Q'_{A},\, A=1,\ldots,N$, which force the system to satisfy the non-holonomic conditions (\ref{nholco}). Equation~(\ref{Lag0}) is then replaced by the following equation,
\beq
\label{Lag3}
\sum_{A=1}^{N} \left \{\frac{d}{dt}\left (\frac{\partial T}{\partial  \dot{q}^{A}} \right ) - \frac{\partial T}{\partial q^{A}} - Q_{A} - Q'_{A} \right \} \delta q^{A} = 0,
\eeq
In Eqn.\ (\ref{Lag3})  the virtual displacements $\delta q^{A}, A = 1, \ldots, N$,  can now be regarded as independent.  Thus one obtains  the equations of motion,
\beq
\label{Lagnh}
\frac{d}{dt}\left (\frac{\partial T}{\partial  \dot{q}^{A}} \right ) - \frac{\partial T}{\partial q^{A}} =Q_{A} + Q'_{A}, \,A = 1, \ldots, N.
\eeq
The forces of constraint, $Q'_{A}, A = 1, \ldots, N$, are {\em a priori} unknown, but they are such that, in any instantaneous displacement $\delta q^{A}, A = 1,\ldots, N$, consistent with the constraints (\ref{nholco}), they do no work.  The non-holonomic constraints (\ref{nholco})  imply the following conditions on the possible instantaneous displacements   $\delta q^{A}, A = 1,\ldots, N$,
\beq
\label{cdelta}
\sum_{A=1}^{N}a^{i}_{A}(q) \delta q^{A} = 0,\,  i = 1,..., m < N.
\eeq
For any instantaneous displacements $\delta q^{A}, A = 1,\ldots, N$, which satisfy the conditions (\ref{cdelta}), the work $\delta W'$ done by the constraint forces $Q'_{A}, A=1,\ldots,N$ is equal to zero, 
{\em i.e.}
\beq
\label{workc}
\delta W' \equiv \sum_{A=1}^{N} Q'_{A}\delta q^{A} = 0.
\eeq
The conditions (\ref{cdelta}) and (\ref{workc})  together imply that
\beq
\label{cforce}
Q'_{A} = \sum_{i=1}^{m} \lambda_{i} a^{i}_{A}(q),\, A = 1,\ldots, N,
\eeq
where the quantities $\lambda_{i}, i = 1,\ldots, m$, are time-dependent parameters.  Eqns.\ (\ref{Lagnh}) have  thus been reduced to the following form,
\beq
\label{Lagnh2}
\frac{d}{dt}\left (\frac{\partial T}{\partial  \dot{q}^{A}} \right ) - \frac{\partial T}{\partial q^{A}} =Q_{A} + 
\sum_{i=1}^{m} \lambda_{i} a^{i}_{A}(q), \,A = 1, \ldots, N.
\eeq
These $N$  equations of motion are  consequences of the principle of  d'Alembert. One should still add the $m$ equations of constraint (\ref{nholco}) to the equations of motion above. There are thus altogether  $N+m$ equations for the determination of $N+m$ quantities $q^{A}(t), A = 1,\ldots,N$, and $\lambda_{i}(t), i = 1, \ldots, m$, when appropriate initial conditions for the quantities $q^{1},...,q^{N}$ and $\dot{q}^{1},...,\dot{q}^{N}$ are given.

One should observe that in the argument above, that it is \underline{not}  required that the constraint equations (\ref{nholco})  be in force under general  variations $q_{j} \rightarrow q_{j} + \delta q_{j}$; the constraints (\ref{nholco}) are \underline{only} imposed on the actual motion of the system. 

It is convenient  to assume that the external applied forces can be derived from a generalized potential. This entails unimportant loss of generality. In what follows we thus assume  the existence of a potential $V$ such that
\beq
\label{GenV}
Q_{A} = - \frac{\partial V}{\partial q^{A}} +  \frac{d}{dt}\left (\frac{\partial V}{\partial  \dot{q}^{A}}\right ), A = 1,...,N.
\eeq
Using the notation
\beq
\label{ell0}
L_{0}(q, \dot{q})  \equiv T - V,
\eeq
we then rewrite the equations  (\ref{Lagnh2})  as follows,
\beq
\label{Lagnh3}
\frac{d}{dt}\left (\frac{\partial L_{0}(q, \dot{q})}{\partial  \dot{q}^{A}}\right ) - \frac{\partial L_{0}(q, \dot{q})}{\partial q^{A}} = \sum_{i=1}^{m} \lambda_{i} a^{i}_{A}(q), \;\,A = 1, \ldots, N.
\eeq
The quantity $L_{0}(q, \dot{q})$ defined above in Eqn.\ (\ref{ell0}) depends on the kinetic energy and on the external applied forces of the system under consideration. If there were no non-holonomic constraints then the quantity $L_{0}(q, \dot{q})$ would be the Lagrangian of the system.  

It should be noted, that from he equations (\ref{Lagnh3}) and the constraints (\ref{nholco}) follows a first integral for the system,
\beq
\label{energyint}
E := \sum_{A=1}^{N} \dot{q}^{A}\, \frac{\partial L_{0}(q, \dot{q})}{\partial  \dot{q}^{A}} - L_{0}(q, \dot{q}),
\eeq
where $E$ is the constant energy of the system.

We now consider the variational action principle discussed above, in which the non-holonomic conditions (\ref{nholco}) are taken into account by means of the multiplication rule in the calculus of variations. Consider the following action functional,
\beq
\label{actionSo}
S_{0} := \int dt\,L_{0}(q, \dot{q}).
\eeq
The action principle referred to previously is simply the requirement that the action functional
(\ref{actionSo}) be stationary when the non-holonomic constraints (\ref{nholco}) are in force throughout a suitable  region {\em D} in configuration space, {\em i.e.} when the conditions
\beq
\label{nholcoA}
\sum_{A=1}^{N} a^{i}_{A}(q)  \dot{q}^{A} = 0,\,  i = 1,..., m < N\,;\; (q^{1},...,q^{N}) \in  D,
\eeq
are in force. Using the multiplication rule in the calculus of variations, the action principle formulated above becomes equivalent to the following free variational problem,
\beq  
\label{varnh3}
\delta \int \,dt \left [L_{0}(q, \dot{q}) - \sum_{i=1}^{m}  \mu_{i} \sum_{A=1}^{N}a^{i}_{A}(q)  \dot{q}^{A} \right ] = 0,
\eeq
which involves m  Lagrange multipliers $ \mu_{i}, i = 1,\ldots, m$. 

The variational equations following from Eqn.\ (\ref{varnh3}) are,
\bea
\label{CC1}
\lefteqn{\frac{d}{dt}\left (\frac{\partial L_{0}(q, \dot{q})}{\partial  \dot{q}^{A}}\right ) - \frac{\partial L_{0}(q, \dot{q})}{\partial q^{A}}   =  }\\
& =  &\sum_{i=1}^{m} \dot{ \mu}_{i} a^{i}_{A}(q)   
+ \sum_{i=1}^{m}  \mu_{i}\sum_{B=1}^{N}\left ( \frac{\partial a^{i}_{A}(q)}{\partial q^{B}} - 
\frac{\partial a^{i}_{B}(q)}{\partial q^{A}} \right ) \dot{q}^{B} = 0,\; A  =  1, \ldots, N. \nonumber
\eea
The N equations (\ref{CC1}) together with the  m conditions (\ref{nholcoA}) are supposed to determine the quantities $q^{1},...,q^{N}$ and the Lagrange multipliers $\mu_{1}, \ldots,  \mu_{m},$ when appropriate initial conditions for $q^{1},...,q^{N}$ and $\dot{q}^{1},...,\dot{q}^{N}$ are given.

The variational equations~(\ref{CC1}) are not identical  to the d'Alembertian equations of motion (\ref{Lagnh3}).  The assumptions underlying the variational equations and the d'Alembertian equations of motion, respectively, are also basically different.  The non-holonomic constraints (\ref{nholco}) are only supposed to be valid for the actual motion when one  applies the principle of d'Alembert to derive the equations of motion, whereas these non-holonomic conditions are supposed to be valid throughout a whole appropriate  region $D$ when one considers the variational problem (\ref{varnh3}), as indicated  in Eqn.\ (\ref{nholcoA}) above. 

It  is readily seen that if the system under consideration is holonomic, {\em i.e.} if the one-forms occurring in Eqn.\ (\ref{cdelta}) are integrable, then the d'Alembertian equations  (\ref{Lagnh3}) and the variational equations (\ref{CC1}) are equivalent.  One reaches the same conclusion when the one-forms in (\ref{cdelta})  can be made integrable by multiplication with integrating factors. We will return to the question of existence of such integrating factors in the next section. The forms (\ref{cdelta}) are integrable if the following   conditions are satisfied,
\beq
\label{intcond}
\frac{\partial a^{i}_{A}(q)}{\partial q^{B}} -  \frac{\partial a^{i}_{B}(q)}{\partial q^{A}} = 0, i = 1, \ldots, m, A, B = 1, \ldots, N.
\eeq
When the conditions (\ref{intcond}) are in force, then the d'Alembertian equations of motion (\ref{Lagnh3})  and the variational equations (\ref{CC1}) become identical  upon a change of notation:
$\dot{\mu}_{i} \rightarrow  \lambda_{i}, i = 1, \ldots, m$.

The variational problem (\ref{varnh3}) is readily recognized to be a variational problem with holonomic constraints when the integrability conditions (\ref{intcond}) are in force, since  then there exist functions 
$\Psi^{i}(q)$ such that
\beq
\label{holonCA}
 a^{i}_{A}(q) = \frac{\partial}{\partial q^{A}} \Psi^{i}(q), \; i = 1, \ldots, m.
\eeq
Here and in what follows we tacitly assume that the configuration space is a contractible manifold.
The non-holonomic constraints (\ref{nholco}) then become equivalent  to the following conditions,
\beq
\label{holoA}
\frac{d}{dt} \Psi^{i}(q) =  0, \; i = 1, \ldots, m,
\eeq
from which  the following holonomic constraints follow,
\beq
\label{holoA2}
 \Psi^{i}(q) =  C^{i},\; i = 1, \ldots, m,
 \eeq
where the quantities $C^{i}$ are constants. From the equations (\ref{holonCA}) and (\ref{holoA2}) follows that the variational problem (\ref{varnh3}) is equivalent to the following problem, with holonomic constraints implemented by means of the multiplication rule,
\beq
\label{varnh3A}
\delta \int \,dt \left [L_{0}(q, \dot{q})  + \sum_{i=1}^{m}  \dot{\mu}_{i} \Psi^{i}(q) \right ] = 0.
\eeq

The use of the  principle of d'Alembert and the variational action principle with non-holonomic constraints give rise to basically different  problems, which also result in different equations of motion. When the constraints are holonomic, both problems yield identical equations of motion,  as was demonstrated above. 

It is legitimate to ask whether the d'Alembertian  equations of motion and the variational equations, respectively, in the non-hoholonomic case nevertheless can have the same solutions in general. It was shown by Pars \cite{Pars} in the case of a specific simple example in three-dimensional configuration space, involving one non-holonomic constraint, that the solutions of the d'Alembertian equations of motion and the corresponding variational equations do not coincide in general. In the next section we consider the case of a general $N$-dimensional system satisfying the  conditions outlined in the Introduction, with  one non-holonomic constraint, and prove that  the general solutions to the d'Alembertian and variational equations, respectively, do not coincide for any $N \geq 3$.  
 

\section{Nonequivalence of the principle of d'Alembert and the variational action principle}

We now consider the d'Alembertian equations of motion  (\ref{Lagnh3})  and the variational equations (\ref{CC1}), respectively, but with only one non-holonomic constraint,
\beq
\label{onenonhol}
\sum_{A=1}^{N} a_{A}(q) \dot{q}^{A} = 0,
\eeq
where the functions $a_{A}(q), A = 1, \ldots, N$, are arbitrary functions of the variables
 $q = (q^{1}, \ldots, q^{N})$, except for the condition that not all  of the quantities $a_{A}(q)$ vanish identically.  Naturally it is also assumed the the functions $a_{A}(q), A = 1, \ldots, N$ satisfy appropriate smoothness conditions. The equations (\ref{Lagnh3})  then  reduce to the following set of d'Alembertian equations of motion,
\beq
\label{onedAeq}
\frac{d}{dt}\left (\frac{\partial L_{0}(q, \dot{q})}{\partial  \dot{q}^{A}}\right ) - \frac{\partial L_{0}(q, \dot{q})}{\partial q^{A}} =  \lambda a_{A}(q), \;\,A = 1, \ldots, N,
\eeq
where $\lambda$ is a time dependent parameter. 

It is convenient to introduce the following notation,
\beq
\label{matrixM}
M_{AB}(q)  :=  \frac{\partial a_{A}(q)}{\partial q^{B}} - \frac{\partial a_{B}(q)}{\partial q^{A}},\; A, B = 1, \ldots, N.
\eeq
The variational equations (\ref{CC1}) reduce to the following set of variational equations in the case of one non-holonomic constraint,  
\beq
\label{oneCC1}
\frac{d}{dt}\left (\frac{\partial L_{0}(q, \dot{q})}{\partial  \dot{q}^{A}}\right ) - \frac{\partial L_{0}(q, \dot{q})}{\partial q^{A}}  =   \dot{ \mu}\, a_{A}(q) + \mu \, \sum_{B=1}^{N}M_{AB}(q) \dot{q}^{B} = 0,\; A  =  1, \ldots, N., 
\eeq
where $\mu$ ia a Lagrange multiplier.

The question is now whether the equations (\ref{onedAeq}) and (\ref{oneCC1}), respectively, can have the same solutions for $q^{1}(t), \ldots, q^{N}(t)$ in general, despite the fact that these equations are not identical. It will be shown below that this is not the case.

Naturally, the functions $L_{0}(q, \dot{q})$ and $a_{A}(q)$ entering into the equations (\ref{onedAeq})
and (\ref{oneCC1}), respectively, will have to satisfy appropriate smoothness conditions in order that these equations may have solutions. We will not enter into a discussion of such smoothness conditions, but rather assume that the equations (\ref{onedAeq}) and (\ref{oneCC1}), respectively, have {\em e.g.} $C^{2}$-solutions in some appropriate time-interval, for given initial values for the coordinates $q^{A}$ and the velocities $\dot{q}^{A}, A = 1, \ldots, N$,
\beq
\label{initq}
\left [q^{A}(t)\right ]_{t=t_{0}} = q_{0}^{A},\; \left [\dot{q}^{A}(t)\right ]_{t=t_{0}} = \dot{q}_{0}^{A},\; A = 1, \ldots, N.
\eeq
The initial values $ q_{0}^{A}$ and  $\dot{q}_{0}^{A}$ at $t = t_{0}$ are free parameters  within an appropriate region of the configuration- and velocity space, except for the restriction
\beq
\label{initrestr}
\sum_{A=1}^{N} a_{A}(q_{0}) \dot{q}_{0}^{A} = 0.
\eeq
The condition (\ref{initrestr})  is a consequence  of  the non-holonomic constraint (\ref{onenonhol}). 

In what follows we use the phrase "general initial value conditions" for the conditions which have been
described  and defined above in Eqns.\  (\ref{initq}) and (\ref{initrestr}).

If the functions $a_{A}(q)$ in Eqn.\ (\ref{onenonhol}) satisfy the conditions
\beq
\label{intcoM}
M_{AB}(q) = 0,\; A, B = 1, \ldots, N,
\eeq
for all $q$ in an appropriate contractible domain $D$, then the condition (\ref{onenonhol}) is integrable, {\em i.e.} there exists a function $\Psi(q) \not\equiv 0$ such that
\beq
\label{holocondA}
0 = \sum_{A=1}^{N} a_{A}(q) \dot{q}^{A} = \frac{d}{dt} \Psi(q(t))  \Leftrightarrow \Psi(q) = C,
\eeq
where $C$ is a constant. In this case the system is in fact holonomic. If the integrability condition (\ref{intcoM}) is not in force, there may exist an integrating factor $\Phi(q) \not\equiv 0$, such that the condition
\beq
\label{rednhol}
\sum_{A=1}^{N} \Phi(q) a_{A}(q) \dot{q}^{A} = 0,
\eeq
is integrable.  In this case  the system is reducible to a system with a holonomic constraint simply by replacing the constraint (\ref{onenonhol}) by the equivalent constraint (\ref{rednhol}), {\em i.e.}
by making the replacement
\beq
\label{redef}
a_{A}(q)  \rightarrow \Phi(q)a_{A}(q)
\eeq
throughout.

We notice in passing that the non-holonomic constraint  (\ref{onenonhol}) in a two-dimensional configuration space is always reducible to an equivalent holonomic constraint by means of an integrating factor.  In what follows we thus only consider  the cases $N = 3, 4, \ldots$.

The necessary and sufficient conditions for the existence of an integrating factor $\Phi(q)$
are the following conditions \cite{Ince}, 
\beq
\label{suffintM}
a_{A}(q)\,M_{BC}(q) + a_{B}(q)\,M_{CA}(q) + a_{C}(q)\,M_{AB}(q) = 0, \;A, B, C = 1, \ldots, N.
\eeq
The integrability conditions formulated above are neatly expressed using exterior derivatives and differential forms, We refer to {\em e.g.} the text-books by Flanders \cite{Flanders} or Fleming \cite{Fleming} for an exposition of the appropriate formalism.

After these preliminaries we will now prove the following:

\noindent
 {\em Theorem of non-equivalence}: If the constraints (\ref{onenonhol}) are truly non-holonomic, {\em i.e.} neither integrable nor reducible to integrable constraints by means of an integrating factor, then the d'Alembertian equations (\ref{onedAeq}) and variational equations (\ref{oneCC1}) do not  have coinciding solutions for  arbitrary general initial value conditions (\ref{initq}) and (\ref{initrestr}). The dimensionality $N$ of configuration space is supposed to be a finite integer, satisfying the condition 
 $N \geq 3$.
 
We  assume that the d'Alembertian equations (\ref{onedAeq})  have unique smooth  ({\em e.g.} $C^{2}$) solutions $(q^{1}(t), \ldots, q^{N}(t))$ in an appropriate time-interval, satisfying the non-holonomic constraint (\ref{onenonhol}) and the general initial value conditions (\ref{initq}) and (\ref{initrestr}).  
We call such solutions general solutions in what follows. 

The method of proof is by {\em reductio ad absurdum}, {\em i.e.} we make the assumption that   the variational equations (\ref{oneCC1})  have solutions which coincide with the  general solutions of the d'Alembertian equations (\ref{onedAeq}), and show that this assumption leads to contradictions. 

Assume that the equations (\ref{oneCC1}) and (\ref{onedAeq}) have coincident solutions. By subtracting Eqn.\ (\ref{oneCC1}) from Eqn.\ (\ref{onedAeq}), one obtains the following equations,
\beq
\label{diffdACC1}
(\lambda - \dot{\mu})\,a_{A} = \mu \; \sum_{B=1}^{N} M_{AB}(q)\,\dot{q}^{B} = 0, \; A = 1, \ldots, N,
\eeq
which  have to be satisfied by the general solutions $(q^{1}, \ldots, q^{N})$ of the d'Alembertian
equations (\ref{onedAeq}). 

Let us first note that one must necessarily have $\mu \not\equiv 0$ in the equations  (\ref{diffdACC1}) above, since otherwise one would have
\beq
\label{exceptA}
\lambda\,a_{A} \equiv 0,\; A =  1. \ldots, N.
\eeq
The condition (\ref{exceptA}) above implies that  $\lambda \equiv 0$, which can not be true in the case of a general non-holonomic constraint. Hence $\mu \not\equiv 0$.

We introduce the notation
\beq
\label{Gamma}
\Gamma := \frac{\lambda - \dot{\mu}}{\mu}.
\eeq
\noindent
The conditions (\ref{diffdACC1}) are then equivalent to the following equations,
\beq
\label{eqnsM}
\sum_{B=1}^{N} M_{AB}(q)\,\dot{q}^{B} = \Gamma\,a_{A}, \; A = 1, \ldots, N,
\eeq
which is a set of $N$ linear algebraic equations in the variables $\dot{q}^{A}, A = 1, \ldots, N$.

The solutions $\dot{q}^{A}, A = 1, \ldots, N$, of the algebraic equations (\ref{eqnsM}) can be obtained in a fairly explicit form, by using standard properties \cite{Greub} of skew (antisymmetric) matrices. Below we give a brief account of the  properties needed in our analysis of the equations (\ref{eqnsM}). A  more detailed account is  given in  Appendix A at the end of this paper.

Let $M \not\equiv 0$ be an arbitrary skew $N\times N$ matrix, with real valued matrix elements $M_{AB}$, labeled by integers $A, B$ in the range $(1, \ldots, N)$, where $N \geq 3$. Since $M$ is skew, its rank is necessarily an even integer $2p \leq N$, Since $M \not\equiv 0$ we must also have  $p \geq 1$. 
Consider the matrix 
$M^{2}$,
\beq
(M^{2})_{AB} := \sum_{C=1}^{N} M_{AC}M_{CB}.
\label{Msquared}
\eeq
The matrix $M^{2}$ is real and symmetric, and hence it has  $N$ eigenvalues  $-\kappa_{\nu}^{2}$, with corresponding ortho-normal eigenvectors $e_{\nu}$,  $\nu = 1, \ldots, N$. The eigenvalues $- \kappa_{\nu}^{2}$ are non-positive, since $M^{2}$ is also negative semidefinite. The numbers $\kappa_{\nu}$ are therefore real, and can be chosen to be non-negative without loss of generality. The non-negative quantities $\kappa_{\nu}, \nu = 1, \ldots, N$, are determined by the roots of the equation
\beq
\label{rootskappa}
\det (M^{2} + \kappa_{\nu}^{2}\, {\bf 1}) = 0.
\eeq
The matrix $M^{2}$ has the same rank as the matrix $M$. Hence,
\beq
\label{pposkappa}
\kappa_{\nu} > 0,\; \nu = 1, \ldots, 2p,
\eeq
and, if $2p < N$,
\beq
\label{nullkappa}
\kappa_{\nu} = 0, \; \nu = 2p+1, \ldots, N.
\eeq

As shown in Appendix A, one can form a set of  ortho-normal basis vectors 
$b_{\nu}, \nu = 1, \ldots, N$, which satisfy the following equations,
\begin{equation}
\label{eq:condMtext}
\sum_{B=1}^{N} M_{AB}b_{2\nu-1, B} = \kappa_{\nu} b_{2\nu, A}; \; \; \sum_{B=1}^{N} M_{AB}b_{2\nu, B} = - \kappa_{\nu} b_{2\nu-1, A},\;\; \nu = 1, \ldots, p,
\end{equation}
and, if $2p < N$,
\begin{equation}
\label{zerobtext}
\sum_{B=1}^{N} M_{AB} b_{\nu, B} = 0, \; \nu = 2p+1, \ldots, N.
\end{equation}
The basis vectors are ortho-normalized in the following inner product,
\beq
\label{orthob}
(b_{\mu}, b_{\nu}) := \sum_{A=1}^{N} b_{\mu, A}\,b_{\nu, A} = \delta_{\mu,\nu} := \left \{ \begin{array}{ll}
1 & \mbox{if $\mu = \nu$}\\
0 & \mbox{otherwise.}
\end{array}
\right .
\eeq
The basis-vectors $b_{\nu}, \nu = 1, \ldots, N$ are  determined by the elements $M_{AB}$ of the matrix $M$. Thus the components $b_{\mu, A}, \mu = 1, \ldots, N;  A = 1, \ldots, N$, are in principle known quantities once the matrix elements $M_{AB}, A, B = 1, \ldots, N$ are given.

We now return to the equation (\ref{eqnsM}), which was shown above to be a necessary consequence of the assumption that Eqns.\  (\ref{onedAeq}) and (\ref{oneCC1}), respectively, have coincident general solutions. 

It is convenient to consider separately the case $\Gamma \equiv 0$ and $\Gamma \not\equiv 0$.

\large
\noindent
{\em The case} $\Gamma \equiv 0$.
\normalsize

When $\Gamma \equiv 0$, Eqns.\  (\ref{eqnsM}) read as follows,
\beq
\label{Gammazero}
\sum_{B=1}^{N} M_{AB}(q)\,\dot{q}^{B} = 0, \; A = 1, \ldots, N,
\eeq

We first consider the case when the matrix $M$ is regular, {\em i.e.} $\det(M_{AB}(q)) \not\equiv 0$. This can happen only if the rank $2p$ of $M$ equals $N$, in which case $N$ is an even integer.  The only solutions of Eqns.\  (\ref{Gammazero}) are then the following,
\beq
\label{zerosol1}
\dot{q}^{A} \equiv 0, \; A= 1, \ldots, N.
\eeq
However, the solutions (\ref{zerosol1}) are not possible, since they are not consistent with the general  initial value conditions (\ref{initq}) and (\ref{initrestr}). 

Let then the rank $2p$ of $M$ be less than $N$. The general solution of Eqns.\  (\ref{eqnsM}) is in this case a linear combination of the $N-2p$ basis vectors $b_{\nu}, \; \nu = 2p+1, \ldots, N$, {\em i.e.},
\beq
\label{zerosol2}
\dot{q}^{A} = \sum_{\nu=2p+1}^{N} \gamma_{\nu}\,b_{\nu, A}, \; A = 1, \ldots, N,
\eeq
where the quantities  $\gamma_{\nu}, \nu = 2p+1, \ldots, N$ are free parameters.

From the orthogonality conditions (\ref{orthob}) follows  then that the solutions (\ref{zerosol2}) are orthogonal to the basis vectors $b_{\mu}$ when $\mu = 1, \ldots, 2p$, {\em i.e.}, 
\beq
\label{condzerosol2}
(b_{\mu}, \dot{q}) = \sum_{A=1}^{N} b_{\mu, A}\,\dot{q}^{A} = 0, \; \mu = 1, \ldots, 2p.
\eeq
The conditions (\ref{condzerosol2}) must in particular also be satisfied for the initial values of $q^{A}$ and $\dot{q}^{A}, A = 1, \ldots, N$, at $t = t_{0}$, {\em i.e.},
\beq
\label{i-cond2}
\sum_{A=1}^{N} \left [b_{\mu, A}\,\dot{q}^{A}\right ]_{t=t_{0}} = 0, \; \mu = 1, \ldots, 2p.
\eeq
Since the number $2p$ is in the range $(2,\ldots, N-1)$, the number of conditions in (\ref{i-cond2}) is  at most $N-1$, but at least $2$.  Thus the solutions (\ref{zerosol2}) are not possible, since they involve at least two  special conditions of the form (\ref{i-cond2}) on the $2N$ initial values $q_{0}^{A}$ and 
$\dot{q}_{0}^{A}, A = 1, \ldots, N$. This is not consistent with the general initial value conditions 
(\ref{initq}) and (\ref{initrestr}).

We have now shown that the case $\Gamma \equiv 0$ leads to contradictions in any $N$-dimensional configuration space of dimension $N \geq 3$. It remains to consider the case $\Gamma \not\equiv 0$.

\large
\noindent
{\em The case} $\Gamma \not\equiv 0$.
\normalsize

We return to the equations (\ref{eqnsM}), where $\Gamma \not \equiv 0$,
\beq
\label{eqnsMGamneq0}
\sum_{B=1}^{N} M_{AB}(q)\,\dot{q}^{B} = \Gamma\,a_{A}; \;  \Gamma \not\equiv 0, \; A = 1, \ldots, N. 
\eeq
We multiply the equations (\ref{eqnsMGamneq0}) above from the left with  the basis vector components $b_{\nu, A}$, where $\nu \leq 2p$, and sum over $A$ in the range $(1, \ldots, N)$.
Using the antisymmetry of the matrix elements $M_{AB}$ and the equations (\ref{eq:condMtext}), one obtains,
\beq
\label{dotqb-comp1}
\kappa_{\nu} (b_{2\nu - 1}, \dot{q}) = \Gamma(b_{2\nu}, a),\;\nu = 1, \ldots, p.
\eeq
and
\beq
\label{dotqb-comp2}
\kappa_{\nu} (b_{2\nu}, \dot{q}) = - \Gamma(b_{2\nu-1}, a),\;\nu = 1, \ldots, p.
\eeq
It is convenient to rewrite the conditions (\ref{dotqb-comp1}) and (\ref{dotqb-comp2}) above as follows,
\beq
\label{dotqbfincond}
(b_{\mu}, \dot{q})  \equiv  \sum_{A=1}^{N} b_{\mu, A}\,\dot{q}^{A}  = \Gamma\, \Theta_{\mu},\;
\mu = 1, \ldots, 2p,
\eeq
where
\beq
\label{knownTheta}
\Theta_{\mu} := \sum_{\nu=1}^{p} \kappa_{\nu}^{-1} [ (b_{2\nu}, a) \delta_{\mu, 2\nu-1} - (b_{2\nu-1}, a) \delta_{\mu, 2\nu} ],\; \mu = 1, \ldots, 2p. 
\eeq
It should be observed that the expressions for the quantities $\Theta_{\mu}$ defined above in Eqn.\ 
(\ref{knownTheta}) contain only quantities which  in principle are known when the components $a_{A}, A = 1, \ldots, N$, occurring in the non-holonomic constraints (\ref{onenonhol}) are given, namely the non-vanishing quantities  $\kappa_{\nu}$ as well as the components of the basis vectors  $b_{2\nu-1}$ and $b_{2\nu}$, for $\nu = 1, \ldots, p$. Consider now the conditions  obtained from the equations (\ref{dotqbfincond}) for any two distinct values of $\mu$, {\em e.g.} $\mu = \alpha$ and $\mu = \beta$. Eliminating the unknown but non-vanishing quantity $\Gamma$ from these equations one obtains the following $p(2p-1)$ conditions,
\beq
\label{lastcond}
(b_{\alpha}, \dot{q})\Theta_{\beta} - (b_{\beta}, \dot{q})\Theta_{\alpha} = 0,\; \alpha, \beta = 1, \ldots, 2p,
\eeq
which hold for all $t$, and hence in particular for the initial time $t_{0}$.

Since $p \geq 1$, there is at least one condition of the form (\ref{lastcond}) involving the initial values for $q^{A}$ and $\dot{q}^{A}, A = 1, \ldots, N$, at $t = t_{0}$. Hence the solutions $\dot{q}^{A}, A = 1, \ldots, N$  of the equations (\ref{eqnsMGamneq0}) are not possible, since they involve at least one special  condition of the form (\ref{lastcond}) on the $2N$ initial values  $q_{0}^{A}$ and $\dot{q}_{0}^{A}, A = 1, \ldots, N$, and this is not consistent with the general initial value conditions (\ref{initq}) and (\ref{initrestr}).

We have now shown that the also the case $\Gamma \not\equiv 0$ leads to contradictions in any $N$-dimensional configuration space of dimension $N \geq 3$. 

\large
\noindent
{\em Conclusion of the proof.}
\normalsize

We have shown that the equations (\ref{eqnsM}), which are an inescapable consequence of the assumption that the d'Alembertian equations of motion (\ref{onedAeq}), and the variational equations (\ref{oneCC1}), respectively, have coincident general solutions, are in contradiction with the requirement that  the initial values $ q_{0}^{A}$ and  $\dot{q}_{0}^{A}$ at $t = t_{0}$ for general solutions must be considered as free parameters in an appropriate region, except for the restriction implied by the given non-holonomic constraint (\ref{onenonhol}) at the initial time $t_{0}$. We have arrived at this contradiction when the constraint  (\ref{onenonhol})  is truly non-holonomic.  The contradiction proves the assertion that the d'Alembertian equations (\ref{onedAeq}) and variational equations (\ref{oneCC1}) do not  have coinciding general solutions  in any $N$-dimensional configuration space with $N \geq 3$, if the constraints  (\ref{onenonhol}) are truly non-holonomic, {\em i.e.} neither integrable nor reducible to integrable constraints by means of an integrating factor. This concludes the proof.

It was demonstrated  in {\em Sec. 2} above, that the d'Alembertian equations of motion (\ref{Lagnh3}) and the variational equations (\ref{CC1}) become identical after a simple change of notation, if the constraints (\ref{nholco}) are holonomic, or reducible to a set of holonomic constraints. When the equations in question are identical, they naturally have identical solutions, as long as  the solutions exist. This conclusions holds  {\em a fortiori} in the case of one constraint. So, we  have in fact proved that the d'Alembertian equations of motion  (\ref{onedAeq}) and the variational equations (\ref{oneCC1}), respectively, have coincident general solutions in the case of one constraint  of the form 
(\ref{onenonhol})  if and only if the constraint in question is either holonomic or reducible to a holonomic constraint by means of an integrating factor. 
 
We finally make the observation that it was sufficient for the proof of the theorem above,  to confront the consequences of the equations (\ref{Gammazero}) and (\ref{eqnsMGamneq0}), respectively, with the requirement that the general solution $(q^{1}(t), \ldots, q^{N}(t))$ of the d'Alembertian equations of motion (\ref{onedAeq}) must be specified by the general initial value conditions (\ref{initq}) and 
(\ref{initrestr}), which involve $2N-1$ initial values as free parameters in an appropriate region. This requirement by no means exhausts the consequences of the equations (\ref{Gammazero}) or 
(\ref{eqnsMGamneq0}), respectively.

\section{Summary and discussion}

It has been proved in this paper that the d'Alembertian equations of motion for a fairly general system with non-holonomic constraints, in three or more dimensions, are not equivalent to the corresponding variational equations with the constraints implemented by the multiplication rule in the calculus of variations. The non-equivalence means two things, namely that the equations in question are not identical in form, and also that they do not have coincident general solutions. The non-coincidence of the general solutions of the d'Alembertian and variational equations, respectively, has been demonstrated in a special case in three dimensions in a paper by Pars \cite{Pars}, but has not until now, to the best of our knowledge, been demonstrated for general systems in $N$ dimensions with 
$N \geq 3$.

The action principle, which has been proposed several times in the literature  for systems with non-holonomic constraints, in analogy with the valid action principle for systems with holonomic constraints, is thus not consistent with the principle of d'Alembert.  It is possible to resolve this problem of inconsistency by dismissing the principle of d'Alembert as a valid principle for systems with non-holonomic constraints. This possibility has been explicitly rejected for good reasons in the papers by {\em e.g.} by Jeffreys \cite{Jeffreys} and Pars \cite{Pars}, which were referred to previously. Also {\em e.g.} in the classic text by Whittaker \cite{Whittaker},  the principle of d'Alembert is considered to be 
fundamental.

It is possible that the d'Alembertian equations of motion in specific cases may have solutions which coincide with the solutions of the corresponding variational equations for some special initial values. In such cases the solutions in question do naturally also follow from the action principle involving the multiplication  rule in the calculus of variation. However this action principle would in such a case not be universal, but would be tied to particular cases with special initial values.

It is legitimate to ask whether there may nevertheless be {\em some} action principle governing systems with non-holonomic constraints.  The analysis of this question can only use the equations of motion in the non-holonomic case, and will then have to rely on such results which are available  for the so-called inverse problem in the calculus of variations \cite{Douglas}, \cite{Sarlet}.  Thus any variational principle which might be found in this manner for some specific system would not be universal. 

In a previous paper \cite{CCTR} we have considered the existence of a variational principle for the simple three-dimensional example with one non-holonomic constraint considered by Pars \cite{Pars}, and have shown that there is indeed an action principle in that case, in the sense that  there is a related system in two dimensions which can be formulated in terms of an action principle involving a  Lagrangian and also a Hamiltonian. This related system  yields the equations of motion for the three-dimensional example.  The Lagrangian for this case is non-unique in a non-trivial way, however.

\vfill\eject

\setcounter{section}{0}
\setcounter{equation}{0}
\renewcommand{\thesection}{Appendix \Alph{section}.}
\renewcommand{\theequation}{A.\arabic{equation}}

\section{Properties of skew matrices}

In this Appendix we give a brief account of those properties of skew (antisymmetric) matrices that are needed  in Section 3 of this paper.  For a detailed account of the material presented here, we refer to a text-book by Greub \cite{Greub}.

Let $M$ be an arbitrary skew  $N\times N$ matrix, with real-valued matrix
elements $M_{AB}$, labeled by integers $A, B$ in the range $(1, \ldots, N)$,  where $N \geq 3$. Consider the matrix $M^{2}$,
\begin{equation}
(M^{2})_{AB} := \sum_{C=1}^{N} M_{AC}M_{CB}.
\label{eq:A2}
\end{equation}
From the anti-symmetry of the matrix $M$ follows that the matrix $M^{2}$ is symmetric and negative semi-definite.  Hence there exists  a set of ortho-normal eigenvectors $\left \{e_{\nu}\right \}, N = 1,\ldots, N$, which solve the following eigenvalue equations,
\begin{equation}
\sum_{B=1}^{N} (M^{2})_{AB} e_{\nu, B} = - \kappa^{2}_{\nu}\;e_{\nu, A}, \;\nu = 1,...,N\,; A = 1, 2,\dots, N,
\label{eq:A3}
\end{equation}
where the quantities  $\kappa_{\nu}$ are real numbers, which without loss of generality  can be assumed to be non-negative.  The eigenvalues and eigenvectors are labeled by Greek letters such as 
$\mu, \nu, \ldots$,  and the single indices  $A, B, \ldots$ label the components of the eigenvectors, or, more generally, the components of any vectors in the domain of M. The inner product $(u, v)$ of any two such vectors $u$ and $v$ is taken to be the following Euclidean inner product,
\begin{equation}
\label{eq:innprodA}
(u, v) := \sum_{A=1}^{N} u_{A}v_{A}.
\end{equation}
Since one uses an Euclidean inner product it is not necessary to distinguish between upper and lower indices of the components of vectors in the domain of $M$. We will use the inner product (\ref{eq:innprodA}) also if one or both of the vector components are labeled by upper indices.
The eigenvectors $e_{\nu}, \nu = 1, \ldots, N$,  are ortho-normalized, 
\begin{equation}
\label{orthoeA}
(e_{\mu}, e_{\nu}) = \delta_{\mu \nu} = \left \{ \begin{array}{ll}
1 & \mbox{if $\mu = \nu$}\\
0 & \mbox{otherwise.}
\end{array}
\right .
\end{equation}

The rank of a skew matrix is always an even non-negative integer. The rank of $M^{2}$ is the same as the rank of M. Let this rank be denoted by $2p$ in what follows.  Thus there are $2p$ eigenvalues 
$- \kappa^{2}_{\nu}$ in the equations (\ref{eq:A3}) which are different from zero, {\em i.e.}
\begin{equation}
\kappa_{\nu} > 0, \; \nu = 1,...,2p.
\label{eq:A4}
\end{equation}
If $2p < N$ there are  $N-2p$  remaining eigenvalues which are zero, {\em i.e.},
\begin{equation}
\kappa_{\nu} = 0, \; \nu = 2p+1,...,N.
\label{eq:A5}
\end{equation}

\noindent
Define a set of $N$ basis vectors $b_{\nu}$ as follows,
\bea
\label{eq:basis}
b_{2\mu -1, A}  & := & e_{\mu, A},\; \mu = 1, \ldots, p, \nonumber \\ 
b_{2\mu, A} & := & \kappa_{\mu}^{-1}\sum_{B=1}^{N}M_{AB}e_{\mu, B}, \; \;\mu = 1,\ldots p, \\
A & = & 1, \ldots, N, \nonumber
\eea
and, when $2p < N$,
\begin{equation}
\label{zerobasA}
b_{\nu, A} := e_{\nu, A}, \; \nu = 2p+1, \ldots, N; \; A = 1, \ldots, N.
\end{equation}
The basis-vectors $b_{\mu}, \; \mu = 1, \ldots, N$ are ortho-normal,  
\begin{equation}
\label{orthoA}
(b_{\mu}, b_{\nu})  = \delta_{\mu \nu}.
\end{equation}

\noindent
We finally note that the equations (\ref{eq:basis}) and (\ref{zerobasA}) imply the following conditions,
\begin{equation}
\label{eq:condMA}
\sum_{B=1}^{N} M_{AB}b_{2\nu-1, B} = \kappa_{\nu} b_{2\nu, A}; \; \; \sum_{B=1}^{N} M_{AB}b_{2\nu, B} = - \kappa_{\nu} b_{2\nu-1, A},\;\; \nu = 1, \ldots, p,
\end{equation}
and, if $2p < N$,
\begin{equation}
\label{zerobA}
\sum_{B=1}^{N} M_{AB} b_{\nu, B} = 0, \; \nu = 2p+1, \ldots, N.
\end{equation}
In the basis $\{b_{\mu} \}_{1}^{N}$, the matrix $M$ has the following so-called normal form,
\begin{eqnarray}
\label{eq:A1}
 M \sim \pmatrix{0&\kappa_{1}&\ldots&\ldots&\ldots&\ldots&\ldots&\ldots\cr
-\kappa_{1}&0&\ldots&\ldots&\ldots&\ldots&\ldots&\ldots\cr
\vdots&\vdots&\ddots&\vdots&\vdots&\vdots&\vdots&\vdots\cr
\ldots&\ldots&\ldots&0&\kappa_{p}&\ldots&\ldots&\ldots\cr
\ldots&\ldots&\ldots&-\kappa_{p}&0&\ldots&\ldots&\ldots\cr
\ldots&\ldots&\ldots&\ldots&\ldots&0&\ldots&\ldots\cr
\ldots&\ldots&\ldots&\ldots&\ldots&\ldots&\ddots&\ldots\cr
\ldots&\ldots&\ldots&\ldots&\ldots&\ldots&\ldots&0\cr},
\end{eqnarray}
which is an immediate consequence of the equations (\ref{eq:condMA}) and (\ref{zerobA}) above.

Any vector $v$ in the domain of $M$ can be expanded in the basis $\{b_{\mu} \}_{1}^{N}$.
It is convenient  to write such an expansion as follows,
\begin{equation}
\label{expansion}
v = \sum_{\mu=1}^{p} \left ( \alpha_{\mu} b_{2\mu -1} + \beta_{\mu} b_{2\mu} \right ) + \sum_{\mu=2p+1}^{N} \gamma_{\mu} b_{\mu},
\end{equation}
with the understanding that the sum $\sum_{\mu=2p+1}^{N}(\ldots)$ is absent if $2p = N$. The quantities
$\alpha_{\mu}$, $\beta_{\mu}$ and $\gamma_{\mu}$ are readily obtained with the aid of the ortho-normality conditions (\ref{orthoA}),
\begin{equation}
\label{compA}
\alpha_{\nu} = (b_{2\nu-1}, v),\;\;\; \beta_{\nu} = (b_{2\nu}, v);\; \nu = 1, \ldots, p,
\end{equation}
and, when $2p < N$,
\begin{equation}
\label{zercompA}
\gamma_{\nu} = (b_{\nu}, v);\; \nu = 2p+1, \ldots, N.
\end{equation}

We will make frequent use of all the equations (\ref{eq:condMA} -- \ref{zerobA}) and 
(\ref{expansion} -- \ref{zercompA}) above in the main body of the text.

\vfill\eject


\begin{thebibliography}{10}
\bibitem{Flannery} M. R. Flannery, "The enigma of non-holonomic constraints",  Am. J. Phys. {\bf 73}, 265 - 272 (2005).
\bibitem{Mikhlin} S. G. Mikhlin, {\em Mathematical Physics, an Advanced Course}, North-Holland, Amsterdam, 1970.
\bibitem {Hertz} H. Hertz, {\em Principles of Mechanics}, Macmillan, New York, 1896. 
\bibitem{Holder} O. H{\"o}lder, "Ueber die Prinzipien von Hamilton und Maupertuis", G{\"o}tt. Nachrichten, 122 - 157 (1896).
\bibitem{Jeffreys} H. Jeffreys, "What is Hamilton's principle?", Quart. Journ. Mech. and Applied Math., {\bf 7}, 335 - 337 (1954).
\bibitem{Pars} L. A. Pars, "Variation principles in dynamics", Quart. Journ. Mech. and Applied Math., 
{\bf 7}, 338 - 351 (1954). 
\bibitem{Berezin} F. A. Berezin,  "Hamiltonian Formalism in the General Lagrange Problem" (in Russian), Uspehi Mat. Nauk  {\bf 29}, No. 3 (177), 183 - 184 (1974).
\bibitem{Goldstein59} H. Goldstein, {\em Classical Mechanics}, Addison-Wesley Publ. Co., Inc. Reading, Massachusetts, 1959, 6th printing. 
\bibitem{Whittaker} E. Whittaker, {\em A Treatise on the Analytical Dynamics of Particles and Rigid Bodies}, Cambridge  University Press, Cambridge, 1961, 4th ed. 
\bibitem{Ince} E. L. Ince, {\em Ordinary Differential Equations}, Dover Publications, Inc., New York, 1956.
\bibitem{Flanders} H. Flanders, {\em Differential Forms with Applications to the Physical Sciences},  Dover Publications, Inc., New York, 1989.
\bibitem{Fleming} W. Fleming, {\em Functions of Several Variables}, 2nd ed., Springer-Verlag, New York, 1977.
\bibitem {Greub} W. H. Greub, {\em Linear Algebra}, Springer Verlag, Berlin Heidelberg New York, 1967.  
\bibitem{Douglas} J. Douglas, "Solution of the inverse problem of the calculus of variations", Trans. Amer. Math. Soc. {\bf 50}, 71 - 128 (1940).
\bibitem{Sarlet} W. Sarlet, G.Thompson and G. E. Prince, "The inverse problem of the calculus of variations: The use of geometrical calculus in Douglas's analysis", Trans. Amer. Math. Soc. 354, 2897-2919  (2002).
\bibitem{CCTR} C. Cronstr{\"o}m, T. Raita, "On the existence of Hamiltonians for non-holonomic systems", Revised version of Helsinki Institute of Physics Preprint HIP-2007-63/TH; arXiv:0711.3997v2 [physics.ed-ph]. 

 



\end{thebibliography}
\end{document}